\documentclass[sigconf]{acmart} 

\usepackage{booktabs} 

\setcopyright{acmcopyright}
\usepackage[english]{babel}
\usepackage{subcaption}
\usepackage[inline]{enumitem}
\usepackage{amsmath}
\usepackage{stmaryrd}
\usepackage{graphicx}
\usepackage{subcaption}
\usepackage{multirow}
\usepackage[utf8]{inputenc}
\usepackage{color,soul}

\setul{0.5ex}{0.3ex}

\def\ignore#1{}

\copyrightyear{2020}
\acmYear{2020}
\setcopyright{iw3c2w3}
\acmConference[WWW '20]{Proceedings of The Web Conference 2020}{April 20--24, 2020}{Taipei, Taiwan}
\acmBooktitle{Proceedings of The Web Conference 2020 (WWW '20), April 20--24, 2020, Taipei, Taiwan}
\acmPrice{}
\acmDOI{10.1145/3366423.3380011}
\acmISBN{978-1-4503-7023-3/20/04}

\begin{document}

\title{IART: Intent-aware Response Ranking with Transformers in Information-seeking Conversation Systems}

\author{Liu Yang$^{1}$ \quad Minghui Qiu$^2$ \quad Chen Qu $^1$ \quad Cen Chen$^3$ \quad  Jiafeng Guo$^4$ \quad Yongfeng Zhang$^5$ \quad \quad \quad W. Bruce Croft$^1$ \quad Haiqing Chen$^2$}

\affiliation{%
	\institution{
	$^1$ Center for Intelligent Information Retrieval, University of Massachusetts Amherst \quad
	$^2$ Alibaba Group  \\ 
	$^3$ Ant Financial Services Group \quad
	$^4$ Institute of Computing Technology, Chinese Academy of Sciences \\ 
	$^5$ Dept. of Computer Science, Rutgers University}
}
\email{{lyang, chenqu, croft}@cs.umass.edu, {minghui.qmh,haiqing.chenhq}@alibaba-inc.com, chencen.cc@antfin.com}
\email{guojiafeng@ict.ac.cn, yongfeng.zhang@rutgers.edu}

\begin{abstract}
	\noindent Personal assistant systems, such as Apple Siri, Google Assistant, Amazon Alexa, and Microsoft Cortana, are becoming ever more widely used. Understanding user intent such as clarification questions, potential answers and user feedback in information-seeking conversations is critical for retrieving good responses. In this paper, we analyze user intent patterns in information-seeking conversations and propose an intent-aware neural response ranking model ``\textbf{IART}'', which refers to ``\textbf{I}ntent-\textbf{A}ware \textbf{R}anking with \textbf{T}ransformers''. IART is built on top of the integration of user intent modeling and language representation learning with the Transformer architecture, which relies entirely on a self-attention mechanism instead of recurrent nets \cite{NIPS2017_Transformers}. It incorporates intent-aware utterance attention to derive an importance weighting scheme of utterances in conversation context with the aim of better conversation history understanding. We conduct extensive experiments with three information-seeking conversation data sets including both standard benchmarks and commercial data. Our proposed model outperforms all baseline methods with respect to a variety of metrics. We also perform case studies and analysis of learned user intent and its impact on response ranking in information-seeking conversations to provide interpretation of results. 
\end{abstract}







\fancyhead{}
\settopmatter{printacmref=false, printfolios=false}


\maketitle

{\fontsize{8pt}{8pt} \selectfont
	\textbf{ACM Reference Format:}\\
	 Liu Yang, Minghui Qiu, Chen Qu, Cen Chen, Jiafeng Guo, Yongfeng Zhang, W. Bruce Croft, Haiqing Chen. 2020. IART: Intent-aware Response Ranking with Transformers in Information-seeking Conversation Systems. In \textit{Proceedings of The Web Conference 2020 (WWW '20), April 20--24, 2020, Taipei, Taiwan.} ACM, New York, NY, USA, 7 pages. \url{https://doi.org/10.1145/3366423.3380011}}

\section{Introduction}
\label{sec:intro}





The recent boom of artificial intelligence has witnessed the emerging and flourishing of many intelligent personal assistant systems, including Amazon Alexa, Apple Siri, Alibaba AliMe, Microsoft Cortana and Google Assistant. This trend has led to an interest in conversational search systems, where users would be able to access information with conversational interactions. 
Existing approaches to building conversational systems include generation-based methods \cite{DBLP:conf/emnlp/RitterCD11,DBLP:conf/acl/ShangLL15}, retrieval-based methods \cite{DBLP:journals/corr/JiLL14,DBLP:conf/sigir/YanSW16,DBLP:conf/sigir/YanZE17}, and hybrid methods~\cite{Song:2018:ERG:3304222.3304379,yang:cikm19}.  
Significant progress has been made on the integration of conversation context by generating reformulated queries with contexts \cite{DBLP:conf/sigir/YanSW16}, enhancing context-response matching with sequential interactions \cite{DBLP:conf/acl/WuWXZL17}, and learning with external knowledge \cite{DBLP:conf/sigir/YangQQGZCHC18}. 
However, much less attention has been paid on the user intent in conversations and how to leverage user intent for response ranking in information-seeking conversations. 


To illustrate user intent in information-seeking conversations, we show an example dialog from the Microsoft Answers Community\footnote{https://answers.microsoft.com} in Table \ref{tab:user_intent_example}. Microsoft Answers Community is a customer support QA forum where users can ask questions relevant to Microsoft products. Agents like Microsoft employees or other experienced users will reply to these questions. There could be multi-turn conversation interactions between users and agents. We define a taxonomy of user intent following previous research \cite{DBLP:conf/sigir/QuYCTZQ18,DBLP:journals/corr/abs-1901-03489}. We can observe that there are diverse user intents such as ``Original Question (OQ)'', ``Information Request (IR)'', ``Potential Answers (PA)'', ``Follow-up Questions (FQ)'', ``Further Details (FD)'', etc. in an information-seeking conversation. Moreover, several transition patterns can happen between different user intent. For example, given a question from the user, an agent could provide a potential answer directly or ask for some information as clarification questions before providing answers. Users will provide further details regarding the information requests from agents. At the beginning of a conversation, the agent would like to greet customers or express gratitude to users before they move on to next steps. Near the end of a conversation, the user may provide positive or negative feedback about answers from agents, or ask a follow-up question to continue the conversation interactions.

\begin{table*}[]
	\small
	\caption{An example dialog to illustrate user intent transition patterns from the Microsoft Answers Community. The user intent ``OQ'', ``IR'', ``PA'', ``FQ'', ``FD'', ``GG'' denote ``Original Question'', ``Information Request'', ``Potential Answer'', ``Follow-up Question'', ``Further Details'', ``Greetings/ Gratitude'' respectively. We highlight some lexical match between utterances and response candidates. This table is better readable in color. }
	\vspace{-0.15in}
	\label{tab:user_intent_example}
	\begin{tabular}{p{1.4cm} | l | p{12cm} | l}
		\hline \hline
		ID & Role  & Utterances                                                                                                                                                                                                 & Intent \\ \hline \hline
		Utterance-1  & User  & Windows downloaded this \setulcolor{red}\ul{update} ``2018-02 Cumulative \setulcolor{red}\ul{Update} for Windows 10 ......'' But during the \setulcolor{green}\ul{restart} it says ``we couldn't complete the \setulcolor{red}\ul{update}, undoing changes''. So what can I do to stop this?  Thanks                                & \textbf{OQ}     \\ \hline
		Utterance-2  & Agent & Is there any other pending \setulcolor{red}\ul{updates}? Try Download \setulcolor{blue} \ul{troubleshooter} for Win 10.                                                                                                                                & \textbf{IR/ PA}  \\ \hline
		Utterance-3  & User  & Yes, pending \setulcolor{red}\ul{updates} the same one. I already used the built in \setulcolor{blue}\ul{troubleshooter}, it did fix some 3 issues, but doing a \setulcolor{green}\ul{restart} the problem persists. Can I stop \setulcolor{red}\ul{updates} from installing this particular one? Thanks. & \textbf{PA/ FQ}  \\ \hline
		Utterance-4  & User  & Not sure if related but I just saw that Malicious Software Removal of March did not install ...... & \textbf{FD}     \\ \hline  \hline
		Response-1 (\textbf{\textit{Correct}})  & Agent & Try run \setulcolor{blue}\ul{troubleshooter} and then \setulcolor{green}\ul{restart} your PC. If problem persist, open start and search for Feedback and open Feedback Hub app and report this issue.                                                   & \textbf{PA}     \\ \hline
		Response-2 (\textbf{\textit{Wrong}}) & Agent  &  Glad to know that you fixed the issue, and as I said downloading the ``Show or hide \setulcolor{red}\ul{updates}'' \setulcolor{blue}\ul{troubleshooter} and \setulcolor{green}\ul{restarting} the PC will help you. Thank you for asking questions and providing feedback here!    
		& \textbf{GG}  \\ \hline \hline
	\end{tabular}
\end{table*}


Such user intent patterns can be helpful for conversation models to select good responses due to the following reasons:
(1) The intent sequence in conversation context utterances can provide additional signals to promote response candidates with correct intent and demote response candidates with wrong intent. For example, in Table \ref{tab:user_intent_example}, given the intent sequence \textbf{[OQ]} $\rightarrow$ \textbf{[IR/ PA]}  $\rightarrow$	\textbf{[PA/ FQ]} $\rightarrow$ \textbf{[FD]}, we know that the user is still expecting an answer to solve her question. Although both Response-1 and Response-2 show some lexical and semantic similarities with context utterances, only Response-1 has the intent ``Potential Answers'' (PA). In this case, the model should have the capability to promote the rank of Response-1 and demote Response-2. 
(2) Intent information can help the model to derive an importance weighting scheme over context utterances with attention mechanisms. In the given example dialog in Table \ref{tab:user_intent_example}, the model should learn to assign larger weights to utterances on question descriptions (OQ and FQ) and further details (FD) in order to address the information need of the user. 

Most existing neural conversation models do not explicitly model user intent in conversations. More research needs to be done to understand the role of user intent in response retrieval and to develop effective models for intent-aware response ranking in information-seeking conversations, which is exactly the goal of this paper. There is some existing related work from the Dialog System Technology Challenge (formerly the Dialog State Tracking Challenge, DSTC)\footnote{\url{https://www.microsoft.com/en-us/research/event/dialog-state-tracking-challenge/}}. Many DSTC tasks focus on goal oriented conversations like restaurant reservation. These tasks are typically tackled with slot filling \cite{Zhang:2016:JMI:3060832.3061040,DBLP:journals/csl/HoriPHHBITTYK19}, which is not applicable to information-seeking conversations because of the diversity of information needs. Recently in DSTC7 of 2018,\footnote{\url{http://workshop.colips.org/dstc7/}} an end-to-end response selection challenge has been introduced, which shares similar motivation to our work. However, the evaluation treated response selection as a classification task and there was no explicit modeling of user intent. 




In this paper, we analyze user intent in information-seeking conversations and propose neural ranking models with the integration of user intent modeling. Different user intent types are defined and characterized following previous research \cite{DBLP:conf/sigir/QuYCTZQ18,DBLP:journals/corr/abs-1901-03489}. Then we propose an intent-aware neural ranking model for response retrieval, which is built on top of recent breakthroughs in natural language representation learning with Transformers \cite{NIPS2017_Transformers,DBLP:journals/corr/abs-1810-04805}. We refer to the proposed model as ``\textbf{IART}''\footnote{IART is pronounced as ``art''.}, which is ``\textbf{I}ntent-\textbf{A}ware \textbf{R}anking with \textbf{T}ransformers''. IART incorporates intent-aware utterance attention to derive the importance weighting scheme of utterances in conversation context towards better conversation history understanding. 
We conduct extensive experiments with three information-seeking conversation data sets: \textbf{MSDialog}\footnote{\url{https://ciir.cs.umass.edu/downloads/msdialog/}} \cite{DBLP:conf/sigir/QuYCTZQ18}, Ubuntu Dialog Corpus (\textbf{UDC}) \cite{DBLP:journals/corr/LowePSP15}, and another commercial customer service data from the AliMe assistant \cite{alime-demo} in Alibaba group (\textbf{AliMe}). We compare our methods with various neural ranking models and baseline methods on response selection in multi-turn conversations including the recently proposed Deep Attention Matching Network (DAM) \cite{DBLP:conf/acl/WuLCZDYZL18}. The results show our methods outperform all baselines. We also perform visualization and analysis of learned user intent patterns. 





Our contributions can be summarized as follows: 
(1) We analyze user intent in information-seeking conversations for intent-aware response ranking. To the best of our knowledge, our work is the first to explicitly define and model user intent for response ranking in information-seeking conversations. 
(2) We propose an intent-aware response ranking model with Transformers to utilize user intent information for response ranking. (3) Experimental results with three different conversation data sets show that our methods outperform various baselines. We also perform analysis on learned user intent and ranking examples to provide insights. The code  of our model implementation will be released on GitHub\footnote{\url{https://github.com/yangliuy/Intent-Aware-Ranking-Transformers}}.



%
%



\section{Related Work}
\label{sec:rel}




\textbf{User Intent in Conversations.}
Some previous research studied utterance intent modeling in conversation systems \cite{Stolcke2000Dialogue,DBLP:journals/corr/abs-1710-10609,bhatia2014summarizing,Shiga:2017:MIN:3077136.3080787}. \citet{Stolcke2000Dialogue} performed dialog acts classification with a statistical approach on the SwitchBoard corpus, which consists of human-human chit chats conversations. In this paper, we explore how to combine utterance intent modeling with response ranking in conversations, so that the learned user intent of context utterances and response candidates can help the model select better responses in information-seeking conversations.  



\textbf{Conversational Search.}
Our research is relevant to conversational search \cite{radlinski2017theoretical,zhang2018towards,thomas2017misc,DBLP:journals/corr/YangZZGC17}, which has received significant attention recently. Radlinski and Craswell described the basic features of conversational search systems \cite{radlinski2017theoretical}. Zhang et al. \cite{zhang2018towards} introduced the System Ask User Respond (SAUR) paradigm for conversational search and recommendation. In addition to conversational search models, researchers have also studied the medium of conversational search \cite{spina2017extracting,trippas2015towards}. Our research targets at the response ranking of information-seeking conversations, with Transformer based ranking models and the integration of user intent modeling.




\textbf{Neural Conversational Models.}
There is growing interest in research about conversation response generation and ranking with deep learning and reinforcement learning \cite{DBLP:journals/corr/abs-1809-08267}. Existing work includes retrieval-based methods \cite{DBLP:conf/acl/WuWXZL17,DBLP:conf/emnlp/ZhouDWZYTLY16,DBLP:conf/sigir/YanSW16,DBLP:conf/cikm/YanSZW16,DBLP:conf/sigir/YanZE17,DBLP:conf/sigir/YangQQGZCHC18,Tao:2019:MFN:3289600.3290985,DBLP:conf/cikm/QuYQZCCI19}, generation-based methods \cite{DBLP:conf/acl/ShangLL15,DBLP:conf/emnlp/RitterCD11,DBLP:conf/naacl/SordoniGABJMNGD15,DBLP:journals/corr/VinyalsL15,P17-1045,alime-chat}, and hybrid methods~\cite{Song:2018:ERG:3304222.3304379,yang:cikm19}. 
 Our work is a retrieval-based method. \citet{DBLP:conf/acl/WuLCZDYZL18} investigated matching a response with conversation contexts with dependency information learned by Transformers. Our proposed models are also built with Transformer encoders. The main difference between our work and their research is that we  explicitly define and model user intent in conversations. We show that the intent-aware attention mechanism can help improve response ranking in conversations.

\textbf{Neural Ranking Models.}
Recent progress of research on neural approaches to IR has introduced a number of neural ranking models for information retrieval, question answering and conversation response ranking \cite{DBLP:journals/corr/abs-1903-06902}. These models include representation focused models \cite{DBLP:conf/cikm/HuangHGDAH13} and interaction focused models \cite{DBLP:conf/nips/HuLLC14,DBLP:conf/aaai/PangLGXWC16,alime-tl,Guo:2016:DRM:2983323.2983769,Yang:2016:ARS:2983323.2983818}. The neural ranking models proposed in our research adopt Transformers, which are solely based on attention mechanisms, as the encoder to learn representations. 

\section{Our Approach}

\subsection{Problem Formulation}
The research problem of response ranking in information-seeking conversations is defined as follows. We are given an information-seeking conversation data set $\mathcal{D} = \{(\mathcal{U}_i, \mathcal{R}_i,  \mathcal{Y}_i)\}_{i=1}^N$, where $ \mathcal{U}_i = \{u_i^1, u_i^2, \dots, u_i^{t-1} , u_i^t\} $ in which $u_i^t$ is the utterance in the $t$-th turn of the $i$-th dialog. $\mathcal{R}_i$ and $\mathcal{Y}_i$ are a set of response candidates $ \{r_i^1, r_i^2, \dots, r_{i}^k\}_{k=1}^M $ and the corresponding labels $ \{y_i^1, y_i^2, \dots, y_{i}^k\} $, where $y_i^k=1$ denotes $r_i^k$ is a true response for $\mathcal{U}_i$. Otherwise $y_i^k=0$. For user intent information, there are sequence level user intent labels for both dialog context utterances and response candidates $\mathcal{E} = \{ (\mathcal{I}_i^u, \mathcal{I}_i^r)\}_{i=1}^N$, where $\mathcal{I}_i^u$ and $\mathcal{I}_i^r$ are user intent labels for context utterances and response candidates for the $i$-th dialog respectively. Our task is to learn a ranking model $f(\cdot)$ with $\mathcal{D}$ and $\mathcal{E}$. For any given $\mathcal{U}_i$, the model should be able to generate a ranking list for the candidate responses $\mathcal{R}_i$ with $f(\cdot)$. Note that in practice, $\mathcal{E}$ can come from predicted results of user intent classifiers to reduce human annotation costs. In our paper, $\mathcal{E}$ are predicted results of the user intent classifier \cite{DBLP:journals/corr/abs-1901-03489} for MSDialog and Ubuntu Dialog Corpus. For AliMe data, $\mathcal{E}$ is the output of the intention classifier which is a probabilistic distribution over 40 intention scenarios~\cite{alime-demo}. 

\subsection{Method Overview}
\label{sec:method_overview_intent}
In following sections, we describe the proposed method for intent-aware response ranking in information-seeking conversations. The model incorporates intent-aware utterance attention to derive the importance weighting scheme of different context utterances. Given input context utterances and response candidates, we first generate representations from two different perspectives: user intent representations with a trained neural classifier and semantic information encoding with Transformers. Then self-attention and cross-attention matching will be performed over encoded representations from Transformers to extract matching features. These matching features will be weighted by the intent-aware attention mechanism and aggregated into a matching tensor. Finally a two-layer 3D convolutional neural network will distill final representations over the matching tensor and generate the ranking score for the conversation context/ response candidate pair. 


 

\subsection{User Intent Taxonomy}
\label{sec:user_intent_taxonomy}

We use the MSDialog data that consists of technical support dialogs for Microsoft products developed by \citet{DBLP:conf/sigir/QuYCTZQ18}. Over $2,000$ dialogs with $10,020$ utterances were sampled for user intent annotation on Amazon Mechanical Turk.\footnote{\url{https://www.mturk.com/}} A taxonomy of 12 labels presented in Table~\ref{tab:taxonomy} were developed to characterize the user intent in information-seeking conversations. The user intent labels include question related labels (e.g., Original Questions, Clarifying Question, etc.), answer related labels (e.g., Potential Answer, Further Details, etc.), feedback related labels (e.g., Positive Feedback, Negative Feedback) and greeting related labels (e.g., Greetings/ Gratitude), which cover most of the user intent types in information-seeking conversations. In addition to MSDialog, we also consider the Ubuntu Dialog Corpus (UDC) \cite{DBLP:journals/corr/LowePSP15}. User intent annotation is also performed for randomly sampled $4,063$ UDC utterances. More details can be found in~\citet{DBLP:conf/sigir/QuYCTZQ18}.


\begin{table}[h]
	\centering
	\caption{Descriptions of user intent taxonomy.}
	\vspace{-0.15in}
	\footnotesize
	\label{tab:taxonomy}
	\begin{tabular}{@{}l |  l |  p{0.24\textwidth}     }
		\hline \hline
		Code & Label                 & Description                                                                                           \\ \hline \hline
		OQ   & Original Question     & The first question that initiates a QA dialog         \\ \hline
		RQ   & Repeat Question       & Questions repeating a previous question    \\ \hline
		CQ   & Clarifying Question   & Users or agents ask for clarification     \\ \hline
		FD   & Further Details       & Users or agents provide more details    \\ \hline
		FQ   & Follow Up Question    & Follow-up questions about relevant issues \\ \hline
		IR   & Information Request   & Agents ask for information from users                           \\ \hline
		PA   & Potential Answer      & A potential solution to solve the question   \\ \hline
		PF   & Positive Feedback     & Positive feedback for working solutions         \\ \hline
		NF   & Negative Feedback     &Negative feedback for useless solutions     \\ \hline
		GG   & Greetings/Gratitude   & Greet each other or express gratitude      \\ \hline
		JK   & Junk                  & No useful information in the utterance           \\ \hline
		O    & Others                & Utterances that cannot be categorized \\ \hline \hline
	\end{tabular}
\end{table}

\begin{figure*}[th]
	\center
	\includegraphics*[viewport=0mm 0mm 220mm 85mm, scale=0.60]{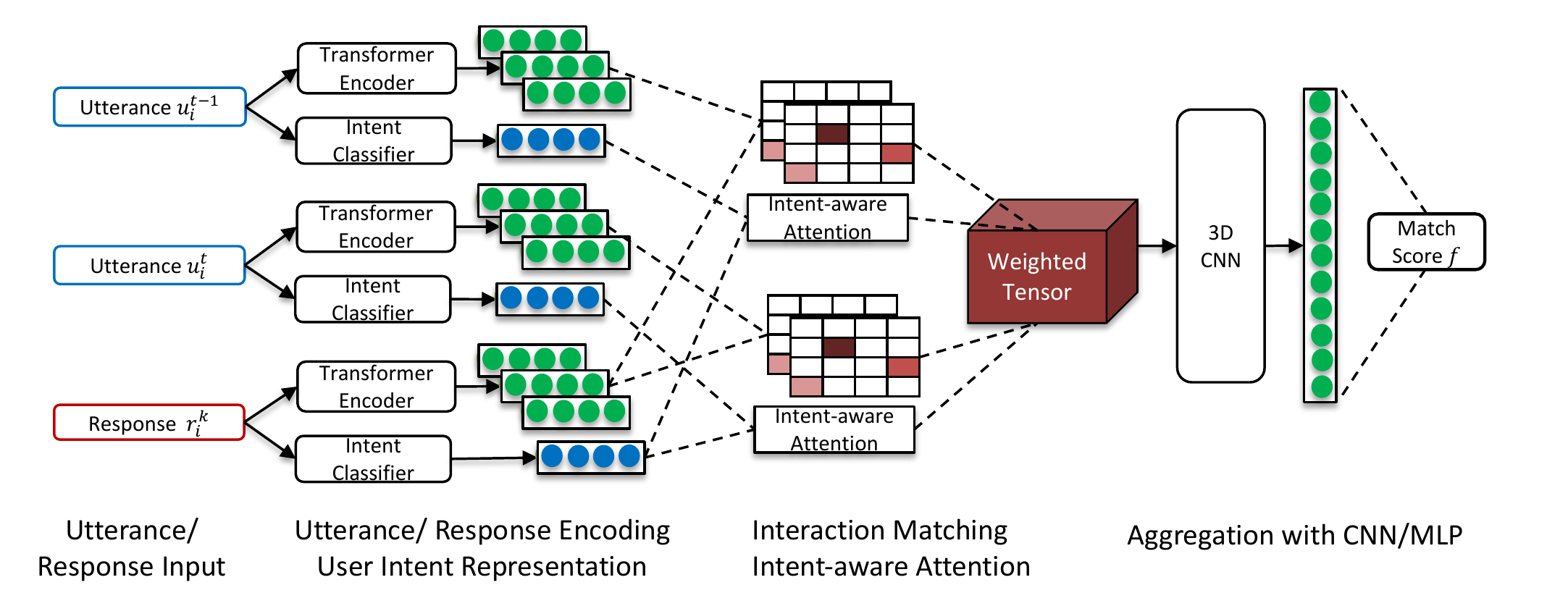}
	\vspace{-0.5cm}
	\caption{The architecture of the IART model for intent-aware conversation response ranking.}\label{fig:iart-attention}
\end{figure*}


\subsection{\textbf{Utterance/ Response Input Representations}}
\label{sec:user_intent_prediction}

Given a response candidate $r_i^{k}$ and an utterance $u_i^t$ in the context $\mathcal{U}_{i}$, we represent the utterance/ response pair from two different perspectives: 1) user intent representation with intent classifiers (Section \ref{sec:user_intent_representaion}); 2) utterance/ response semantic information encoding with Transformers (Section \ref{sec:encoding_with_transformers}).


\subsubsection{\textbf{User Intent Representation}}
\label{sec:user_intent_representaion}

To represent user intent, we adopt the best setting of the neural classifiers CNN-Context-Rep proposed by \citet{DBLP:journals/corr/abs-1901-03489} for user intent classification. Specifically, given sequences of embedding vectors for context utterances and response candidate  $\mathbf{E}(u_i^t)$ and $\mathbf{E}(r_i^{k})$, convolutional filters with the shape $(f, d)$ are applied to a window of $f$ words to produce a new feature $c_i$. This operation is applied to every possible window of words in the utterance $u_i^{t}$ and generates a feature map $\mathbf{c} = \{c_1, c_2, \dots, c_{n-f+1}\}$. Max pooling is applied to select the most salient feature. The model uses multiple filters with varying window sizes to obtain multiple features in different granularity. These features will be concatenated and flattened into an output tensor, which will be projected into a tensor with shape $(l_t,1)$ with a fully connected layer.  $l_t$ is the number of different user intent labels.\footnote{In our experiments for MSDialog and UDC, $l_t = 12$ as presented in Section \ref{sec:user_intent_taxonomy}.} 
\subsubsection{\textbf{Utterance/ Response Encoding and Matching with Transformers}}
\label{sec:encoding_with_transformers}



We adopt the encoder architecture in Transformers \cite{NIPS2017_Transformers} to encode the semantic dependency information in utterance/ response pairs. Transformers are built with Scaled Dot-Product Attention, which performs transformation from a query and a set of key-value pairs to an output. 
Following the design of Transformers, we also add a feed-forward network FFN with ReLU activation over the layer normalized \cite{DBLP:journals/corr/BaKH16} sum of the output $\text{Attention}(\mathcal{Q}, \mathcal{K}, \mathcal{V})$ and the query $\mathcal{Q}$. We refer to this module as the TransformerEncoder module, which will be used as a feature extractor for utterances and responses to capture both the dependency information within words in the same sequence and interactions between words in two different sequences. We consider both self-attention and cross-attention based interaction matching to learn representations for context utterance/ response candidate pairs.

\subsection{\textbf{Intent-aware Attention Mechanism}}
\label{sec:intent_aware_attention}


Given the self-attention/ cross-attention interaction matching matrices for different utterances/ response pairs from a dialog, we first stack them to aggregate them as a 4D matching tensor as follows:

 \begin{footnotesize}
\begin{align}\label{Eqn:4D_Tensor_Agg}
\mathcal{B} = \{ \mathbb{B}_{t,p,q,l} \}_{l_c \times l_u \times l_r \times (2L + 2)}
\end{align}
 \end{footnotesize}
where $l_c,l_u,l_r, L$ are the number of utterance turns in conversation context, number of words in the context utterance, number of words in the response candidate and number of stacked layers in TransformerEncoder. $t,p,q,l$ are indexes along these $4$ dimensions of the matching tensor.

We propose an intent-aware attention mechanism to weight matching representations of different utterance turns in a conversation context, so that the model can learn to attend to different utterance turns in context. The motivation is to incorporate a more flexible way to weight and aggregate matching features of different turns with intent-aware attention. Specifically, let $\mathbf{I}_u^t \in \mathbb{R}^{l_t \times 1}, \mathbf{I}_r^k \in \mathbb{R}^{l_t \times 1}$ denote the intent representation vectors defined in Section \ref{sec:user_intent_representaion} for context utterances and response candidates, we design three different types of intent-aware attention as follows:


\textbf{Dot Product}. We concatenate the two intent representation vectors of the utterance/ response pair, and compute the dot product between the parameter $\mathbf{w}$ and the concatenated vector: $\mathcal{A}_t = \text{softmax}(\exp(\mathbf{w}^T   [\mathbf{I}_u^t, \mathbf{I}_r^k]))$, where $\mathbf{w} \in \mathbb{R}^{2l_t \times 1}$ is a model parameter. 


\textbf{Bilinear}. We compute the bilinear interaction between $\mathbf{I}_u^t$ and $\mathbf{I}_r^k$ and then normalize the result: $\mathcal{A}_t = \text{softmax}( \exp( {\mathbf{I}_u^t}^T \mathbf{w} \mathbf{I}_r^k )) $, where $\mathbf{w} \in \mathbb{R}^{l_t \times l_t}$ is the bilinear interaction matrix to be learned.


\textbf{Outer Product}. We compute the outer product between $\mathbf{I}_u^t$ and $\mathbf{I}_r^k$ and then flatten the result matrix to a feature vector. Finally we project this feature vector into an attention score with a fully connected layer and a softmax function: $ \mathcal{A}_t = \text{softmax} (\exp( \mathbf{w}^T \cdot \text{flat} (\mathbf{I}_u^t \otimes {\mathbf{I}_r^k}^T  )))$, where  flat and $\otimes$ denote the flatten layer which transforms a matrix with shape $(l_t \times l_t)$ into a vector with shape $(l_t^2 \times 1)$ and outer product operation. $\mathbf{w} \in \mathbb{R}^{l_t^2 \times 1}$ is a model parameter. 


Note that the normalization in the softmax function is performed over all utterance turns within a conversation context. Thus the result $\mathcal{A}_t$ is the attention weight corresponding to the $t$-th utterance turn in a conversation context. We also add masks over the padded utterance turns to avoid introducing noise matching feature representations. With the computed attention weights over context utterance turns, we can scale the 4D matching tensor to generate a weighted matching tensor:

 \begin{footnotesize}
\begin{align}\label{Eqn:4D_Tensor_Agg_Weighted}
\widehat{\mathcal{B}} = \{ \mathbb{B}_{t,p,q,l} \cdot \mathcal{A}_t\}_{l_c \times l_u \times l_r \times (2L + 2)}
\end{align}
 \end{footnotesize}
Finally IART adopts a two layer 3D convolution neural network (CNN)\footnote{\url{https://www.tensorflow.org/api_docs/python/tf/nn/conv3d}} to extract important matching features from this weighted matching tensor $\widehat{\mathcal{B}}$. A 3D CNN requires 5D input and filter tensors, as we can add one more input dimension corresponding to the batched training examples over the 4D weighted matching tensor. We compute the final matching score $f(\mathcal{U}_i, r_i^k)$ with a MLP over the flattened output of the 3D CNN. For model training, we compute the cross-entropy loss between the predicted matching scores $f(\mathcal{U}_i, r_i^k)$ and the ground truth matching labels. The parameters of IART are optimized using back-propagation with \textit{Adam} algorithm~\cite{DBLP:journals/corr/KingmaB14}.

\section{Experiments}
\label{sec:exps}
 \subsection{Data Set Description}
 \label{sec:data_desc}

 We evaluated our method with three data sets: Ubuntu Dialog Corpus (UDC), MSDialog, and a commercial data collected from the AliMe assistant at Alibaba group. The statistics of different experimental data sets are shown in Table \ref{tab:exp_data_stat_train_valid_test}. The Ubuntu Dialog Corpus (UDC) \cite{DBLP:journals/corr/LowePSP15} contains multi-turn technical support conversation chat logs on the Ubuntu system. We used the data copy shared by \citet{DBLP:journals/corr/XuLWSW16}. It is also used in several previous related works \cite{DBLP:conf/acl/WuWXZL17,DBLP:conf/acl/WuLCZDYZL18,DBLP:conf/sigir/YangQQGZCHC18}.\footnote{The data can be downloaded from \url{https://www.dropbox.com/s/2fdn26rj6h9bpvl/ubuntu\%20data.zip?dl=0}}
 MSDialog is released from previous related work by \citet{DBLP:conf/sigir/QuYCTZQ18}. It contains QA dialogs on various Microsoft products crawled from the Microsoft Answer community. 
 For the AliMe dataset, it contains the chat logs between customers and the AliMe assistant bot at Alibaba. For each query of the dataset, it contains several response candidates from the chatbot engine which are labeled by a business analyst. The details about these data sets are in~\citet{DBLP:conf/sigir/YangQQGZCHC18}. Note that the proposed model is more on response re-ranking instead of response retrieval in one step.
 

  \begin{table}[]
 	\footnotesize
 	\centering
 	\caption{The statistics of experimental datasets, where C denotes context and R denotes response. \# Cand. per C denotes the number of candidate responses per context. Note that we did not filter any stop words or words with low frequency for computing the average length of contexts or responses.}
 	\vspace{-0.15in}
 	\label{tab:exp_data_stat_train_valid_test}
 	\begin{tabular}{l | p{0.35cm} p{0.35cm} p{0.35cm} | p{0.35cm} p{0.35cm} p{0.35cm} | p{0.3cm} p{0.3cm} p{0.3cm}}
 		\hline \hline
 		Data                               & \multicolumn{3}{c|}{UDC}      & \multicolumn{3}{c|}{MSDialog} & \multicolumn{3}{c}{AliMe} \\ \hline
 		Items                              & Train     & Valid   & Test    & Train     & Valid   & Test    & Train    & Valid  & Test   \\ \hline
 		\# C-R pairs          & 1000k & 500k & 500k & 173k   & 37k  & 35k  & 51k   & 6k  & 6k  \\ \hline
 		\# Cand. per C          & 2         & 10      & 10      & 10        & 10      & 10      & 15       & 15     & 15     \\ \hline
 		\# + Cand. per C & 1         & 1       & 1       & 1         & 1       & 1       & 2.9      & 2.8    & 2.9    \\ \hline
 		Avg \# turns per C & 10.1 & 10.1 & 10.1 & 5.0 & 4.9 & 4.4 & 2.4 & 2.1 & 2.2    \\ \hline
 		Avg \# words per C & 116.8 & 116.3 & 116.7 & 451.3 & 435.2 & 375.1 & 38.3 & 35.3 & 34.2  \\ \hline
 		Avg \# words per R & 22.2 & 22.2 & 22.3 & 106.1 & 107.4 & 105.5 & 4.9 & 4.7 & 4.6 \\ \hline \hline
 	\end{tabular}
 \end{table}

\begin{table*}[]
	\small
	\centering
	\caption{Comparison of different models over Ubuntu Dialog Corpus (UDC), MSDialog, and AliMe data sets.  Numbers in bold font mean the result is better compared with the best baseline DAM. 
		$\dagger$ and $\ddagger$ means statistically significant difference over the best baseline DAM with $p < 0.1$ and $p < 0.05$ measured by the Student's paired t-test respectively. } 
	\vspace{-0.15in}
	\label{tab:exp_res_udc_ms_ec}
	\begin{tabular}{l|l|l|l|l|l|l|l|l|l|l|l|l}
		\hline 	\hline
		Data           & \multicolumn{4}{c|}{UDC}                & \multicolumn{4}{c|}{MSDialog}           & \multicolumn{4}{c}{AliMe}          \\ \hline 
		Methods        & R10@1 & R10@2 & R10@5 & MAP    & R10@1 & R10@2 & R10@5 & MAP    & R10@1 & R10@2 & R10@5 & MAP    \\ \hline
		BM25 \cite{Robertson:1994:SEA:188490.188561}       & 0.5138   & 0.6439   & 0.8206   & 0.6504 & 0.2626   & 0.3933   & 0.6329   & 0.4387 & 0.2371   & 0.4204   & 0.6407   & 0.6392 \\  
		BM25-PRF \cite{DBLP:conf/sigir/YangQQGZCHC18}  & 0.5289   & 0.6554   & 0.8292   & 0.6620 & 0.2652   & 0.3970   & 0.6423   & 0.4419 & 0.2454   & 0.4209   & 0.6510   & 0.6412 \\  \hline
		MV-LSTM \cite{DBLP:conf/aaai/WanLGXPC16}   & 0.4973   & 0.6733   & 0.8936   & 0.6611 & 0.2768   & 0.5000   & 0.8516   & 0.5059 & 0.2480   & 0.4105   & 0.7017   & 0.7734 \\ 
		DRMM \cite{Guo:2016:DRM:2983323.2983769}      & 0.5287   & 0.6773   & 0.8776   & 0.6749 & 0.3507   & 0.5854   & 0.9003   & 0.5704 & 0.2212   & 0.3616   & 0.6575   & 0.7165 \\  
		Duet \cite{Mitra:2017:LMU:3038912.3052579}      & 0.4756   & 0.5592   & 0.8272   & 0.5692 & 0.2934   & 0.5046   & 0.8481   & 0.5158 & 0.2433   & 0.4088   & 0.6870   & 0.7651 \\ 
		DMN-KD \cite{DBLP:conf/sigir/YangQQGZCHC18}    & 0.6443   & 0.7841   & 0.9351   & 0.7655 & 0.4908   & 0.7089   & 0.9304   & 0.6728 & 0.3596   & 0.5122   & 0.7631   & 0.8323 \\  
		DMN-PRF \cite{DBLP:conf/sigir/YangQQGZCHC18}   & 0.6552   & 0.7893   & 0.9343   & 0.7719 & 0.5021   & 0.7122   & 0.9356   & 0.6792 & 0.3601   & 0.5323   & 0.7701   & 0.8435 \\ 
		DAM \cite{DBLP:conf/acl/WuLCZDYZL18}       & 0.7686   & 0.8739   & 0.9697   & 0.8527 & 0.7012   & 0.8527   & 0.9715   & 0.8150 & 0.3819  & 0.5567   &        0.7717  & 0.8452 \\ \hline
		IART\begin{footnotesize}\textit{Dot}\end{footnotesize} & \textbf{0.7703} & \textbf{0.8746} & 0.9688	& \textbf{0.8535} & \textbf{0.7234}$^\ddagger$ & \textbf{0.8650}$^\ddagger$ & \textbf{0.9772}$^\ddagger$ & \textbf{0.8300}$^\ddagger$ &
		\textbf{0.3821} & {0.5547} & \textbf{0.7802}$^\dagger$  &  \textbf{0.8454} \\
		IART\begin{footnotesize}\textit{Outerproduct}\end{footnotesize}  & \textbf{0.7717}$^\ddagger$ & \textbf{0.8766}$^\ddagger$ & 	0.9691 & \textbf{0.8548}$^\ddagger$ & \textbf{0.7212}$^\ddagger$  & \textbf{0.8664}$^\ddagger$  & \textbf{0.9749} & \textbf{0.8289}$^\ddagger$  & \textbf{0.3901}$^\ddagger$ & \textbf{0.5649}$^\ddagger$ & \textbf{0.7812}$^\dagger$ & \textbf{0.8493}$^\dagger$ \\
		IART\begin{footnotesize}\textit{Bilinear}\end{footnotesize} & \textbf{0.7713}$^\ddagger$   & \textbf{0.8747}   & 0.9688   & \textbf{0.8542}$^\dagger$ & \textbf{0.7317}$^\ddagger$   & \textbf{0.8752}$^\ddagger$   & \textbf{0.9792}$^\ddagger$   & \textbf{0.8364}$^\ddagger$ & \textbf{0.3892}$^\dagger$ & \textbf{0.5592}$^\dagger$      & \textbf{0.7801}$^\dagger$ & \textbf{0.8471} \\ \hline \hline
	\end{tabular}
\end{table*}
\begin{table*}[]
	\small
	\centering
	\caption{A case study and examples of Top-1 ranked responses by different methods. $y_i^k$  means the label of a response candidate.} 
	\vspace{-0.15in}
	\label{tab:case_study_intent}
	\begin{tabular}{p{1.8cm}  | p{0.2cm} | p{14.5cm}}
		\hline  \hline
		\multirow{1}{*}{Context} & \multicolumn{2}{p{15cm}}{ \textbf{[User]} Hi, I have the new Outlook which updated a few days ago.  I cannot find how to add senders to my blocked senders list manually. How do I do this on the new Outlook? Thanks    \qquad  \textbf{[Agent]}  Hi,  There are different ways to block senders on Outlook depending on the version of Outlook that you are using. May we know what version of Outlook are you using? \qquad \textbf{[User]} Hi,  I'm using the desktop website beta version.  Thanks. \qquad  \textbf{[Agent]}  Desktop Website beta version? Are you referring to the Outlook Web App or the Windows mail? \qquad  \textbf{[User]} I go to Outlook.com and sign in on there.  } \\  \hline
		\multicolumn{2}{l|}{Context Intent}  & \textbf{[OQ]} $\rightarrow$ \textbf{[IR]}  $\rightarrow$	\textbf{[PA]} $\rightarrow$ \textbf{[IR]}   $\rightarrow$ \textbf{[FD/ OQ]}  \\ \hline
		Method      & $y_i^k$   & Top-1 Ranked Response    \\ \hline
		DAM     & 0    & Thanks for the reply.  Some email domain needs to be manually added to Outlook. However, it's good to know that the issue is resolved from your end.  Should you need further assistance in the future, please do let us know.    \textbf{{[}PF{]}}      \\ \hline
		IART\begin{footnotesize}\textit{Bilinear}\end{footnotesize}  & 1    &  In Outlook Web App ...... to manually block an email address, follow these steps: ......   Let us know how things go.    \textbf{{[}PA{]}}      \\ \hline \hline
	\end{tabular}
\end{table*}

\subsection{Experimental Setup}

\subsubsection{\textbf{Baselines}}
We consider different baselines as follows \footnote{Note that the experimental setup where we compare our method with baselines without user intent modeling is reasonable. User intent modeling should be only added into the treatment instead of baselines for controlled experimental comparison to show the effectiveness of the incorporation of user intent.}: 


\textbf{Traditional retrieval models}: these methods treat the dialog context as the query to retrieve response candidates for response selection. We consider BM25 \cite{Robertson:1994:SEA:188490.188561} as the retrieval model. We also consider BM25-PRF \cite{DBLP:conf/sigir/YangQQGZCHC18}, which matches conversation context with the expanded responses using BM25.

\textbf{Neural ranking models}: we consider several representative neural ranking models: MV-LSTM \cite{DBLP:conf/aaai/WanLGXPC16}, DRMM \cite{Guo:2016:DRM:2983323.2983769} and Duet \cite{Mitra:2017:LMU:3038912.3052579}.
We also consider models based on Deep Matching Networks (DMN) with external knowledge \cite{DBLP:conf/sigir/YangQQGZCHC18}, which incorporate external knowledge with pseudo-relevance feedback (DMN-PRF) and QA correspondence knowledge distillation (DMN-KD). 

\textbf{Deep Attention Matching Network (DAM)}~\cite{DBLP:conf/acl/WuLCZDYZL18}: DAM is a strong baseline method for response ranking in multi-turn conversations with open source code released\footnote{\url{https://github.com/baidu/Dialogue/tree/master/DAM}} until this paper. DAM also represents and matches a response with its multi-turn context using dependency information learned by Transformers. It does not explicitly model user intent in conversations.

For evaluation metrics, we adopted mean average precision (MAP) and $R_n@k$ which is the recall at top $k$ ranked responses from $n$ available candidates for a given conversation context following previous related works \cite{DBLP:conf/acl/WuLCZDYZL18,DBLP:conf/sigir/YangQQGZCHC18,DBLP:conf/acl/WuWXZL17,DBLP:journals/corr/LowePSP15}. 

\subsubsection{\textbf{ Parameter Settings and Implementation Details}.} 
All models are implemented with TensorFlow\footnote{\url{https://www.tensorflow.org/}} and the MatchZoo\footnote{\url{https://github.com/NTMC-Community/MatchZoo}} toolkit. Hyper-parameters are tuned with the validation data. For the hyper-parameter settings of IART, we set the size of the convolution and pooling kernels as $(3,3,3)$. The number of stacked Transformers layers is set as $5$ for UDC and $4$ for MSDialog. The batch size is $128$ for UDC and $32$ for MSDialog. All models are trained on a single Nvidia Titan X GPU. Learning rate is initialized as 1e-3 with exponential decay during training process. The decay steps and decay rate are set as $400$ and $0.9$. The maximum utterance length is $50$ for UDC and $200$ for MSDialog. The maximum number of context utterance turns is set as $9$ for UDC and $6$ for MSDialog. We padded zeros if the number of utterance turns in a context is less than the maximum number of utterance turns. For user intent labels, there are $12$ different types for UDC/ MSDialog, and $40$ different types for AliMe data. For the word embeddings, we trained word embeddings with the Word2Vec tool \cite{DBLP:conf/nips/MikolovSCCD13} with the CBOW model using our training data following previous work \cite{DBLP:conf/acl/WuWXZL17,DBLP:conf/acl/WuLCZDYZL18}. The max skip length between words and the number of negative examples is set as $10$ and $25$. The dimension of word embeddings is $200$. Word embeddings will be initialized by these pre-trained word vectors and updated during the training process.


\subsection{Evaluation Results}


We present evaluation results over different methods in Table \ref{tab:exp_res_udc_ms_ec}. We summarize our observations as follows: (1) On MSDialog, all three variations of IART with dot, outer product and bilinear based intent-aware attention mechanism show significant improvements over all baseline methods, including the recently proposed strong baseline method DAM. On UDC, IART with three different intent-aware attention mechanisms also show improvements under all metrics except for R10@5. With the comparison between the results of DAM and IART, we can find that incorporating user intent modeling and intent-aware attention weighting scheme can help improve the response ranking performance. (2) If we compare three variations of IART, we can find that the bilinear based intent-aware attention mechanism works better for MSDialog and outer product based intent-aware attention mechanism works better for UDC. The overall performances of these three model variations are close to each other.  Overall our proposed model IART shows larger performance improvements on MSDialog. One possible reason is that the intent classifier on MSDialog is more accurate due to the larger annotated training data of MSDialog for user intent prediction and more formal language used in MSDialog, as shown in evaluation results by \citet{DBLP:journals/corr/abs-1901-03489}. (3) On AliMe data, all three variations of IART also show comparable or better results than all baseline methods including the strong baseline DAM. These results on real product data further verify the effectiveness of our proposed methods. 

\subsection{Case Study and User Intent Visualization}

\begin{figure} 
	\centering
	\includegraphics[width=0.30\textwidth]{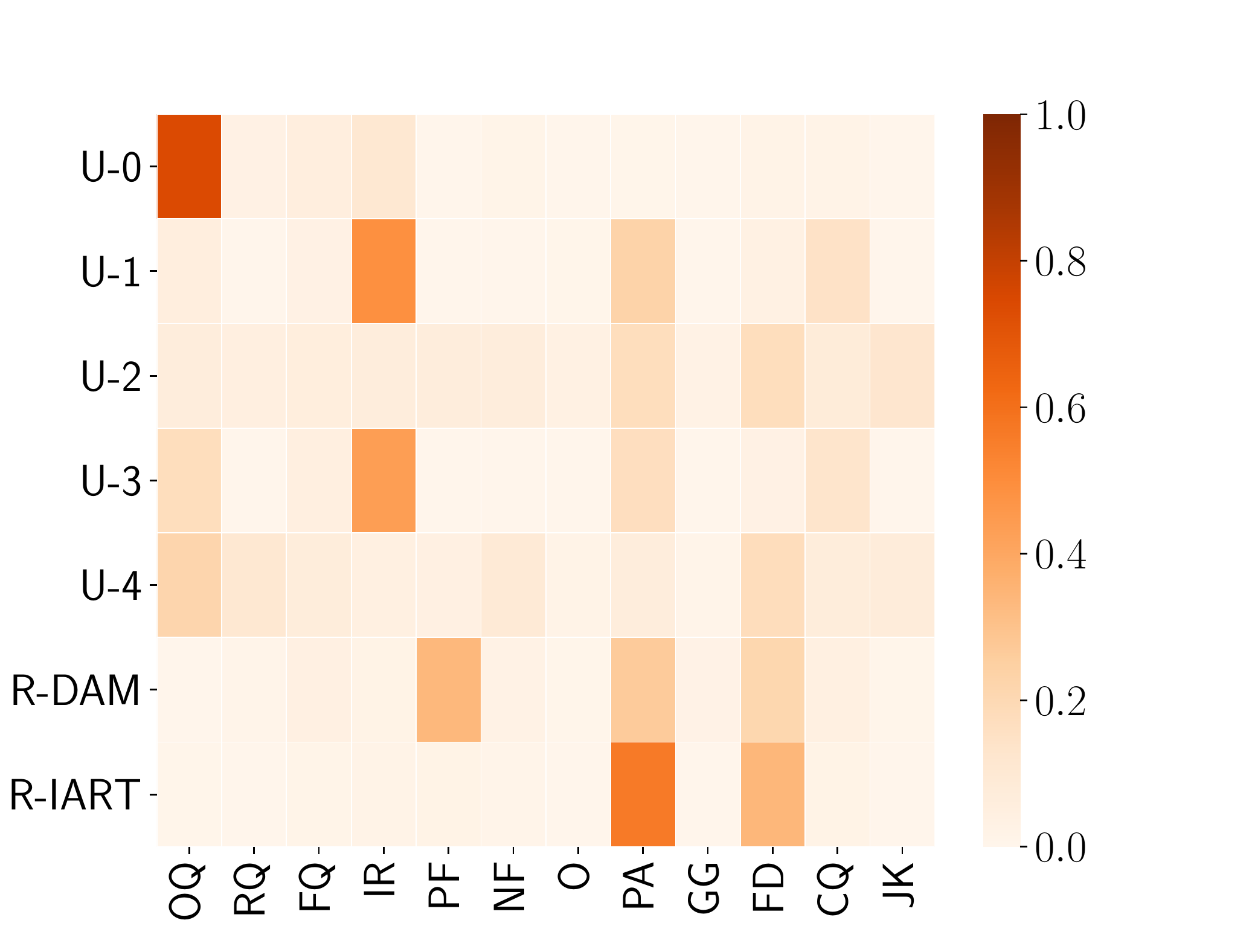}
	\vspace{-0.4cm}
	\caption{Visualization of learned user intent representation of context utterances and returned top-1 ranked response by DAM and IART from the case study in Table \ref{tab:case_study_intent}. U-0 to U-4 denotes the 0-th turn to the 4-th utterance turn in the context. R-DAM and R-IART denotes the top-1 ranked response returned by DAM and IART respectively. Darker spots mean higher predicted probabilities.}\label{fig:user_intent_visual}
\end{figure}

We perform a case study in Table \ref{tab:case_study_intent} on the top ranked responses by different methods including the best baseline DAM and our proposed model IART with bilinear based intent-aware attention mechanism. We show the conversation context utterances and top-1 ranked response by each method. In this example, IART produced the correct top ranked response. We visualized the learned user intent representation of context utterances and returned top-1 ranked response by DAM and IART in Figure \ref{fig:user_intent_visual}. The predicted user intent of conversation utterances is \textbf{[OQ]} $\rightarrow$ \textbf{[IR]}  $\rightarrow$	\textbf{[PA]} $\rightarrow$ \textbf{[IR]}   $\rightarrow$ \textbf{[FD/ OQ]}. The agent performed ``Information Request (IR)'' to confirm whether it is the Outlook Web app or the Windows desktop app. The user confirmed ``Further Details (FD)'' that the problem was related to the Outlook Web app (Outlook.com). Given such a user intent pattern in the conversation context, a reasonable response can be with intent ``Potential Answers (PA)'' on providing potential solutions to the user's question, which is captured by IART due to the integration of user intent modeling. The DAM model, without user intent modeling, failed in such cases and selected a response candidate with ``Positive Feedback (PF)'' intent. The response returned by DAM assumed that ``the issue is resolved'', but actually the user was expecting an answer to her unsolved technical problem. This gives an example and interpretation of why user intent modeling can be helpful for response ranking in conversations.



%
%
%
%

\section{Conclusions}
\label{sec:conclu}

In this paper, we analyze user intent in information-seeking conversations and propose an intent-aware neural ranking model with Transformers. 
We first define and characterize different user intent types, and then propose an intent-aware neural ranking model for response retrieval which incorporates intent-aware utterance attention to derive the importance weighting scheme of different utterances 
to improve conversation history understanding. Our proposed methods outperform all baseline methods 
on three different data sets including both standard benchmarks and commercial data. We also perform case studies and analysis of the learned user intent with their impact on response ranking in information-seeking conversations to provide insights. 





\section{Acknowledgments}
This work was supported in part by the Center for Intelligent Information Retrieval, in part by NSF IIS-1715095, and in part by China Postdoctoral Science Foundation (No. 2019M652038). Any opinions, findings and conclusions or recommendations expressed in this material are those of the authors and do not necessarily reflect those of the sponsor.

\bibliographystyle{ACM-Reference-Format}
\balance
\bibliography{main}

\end{document}